\documentclass[letterpaper]{article}

\usepackage[T1]{fontenc}

\usepackage{geometry}
\usepackage{setspace}
\usepackage{lineno}

\usepackage[style = chem-acs]{biblatex}
\usepackage{graphicx}
\usepackage{float}
\newfloat{scheme}{htbp}{los}
\floatname{scheme}{Scheme}
\floatname{chart}{Chart}
\newfloat{graph}{htbp}{loh}

\usepackage{chemformula} 
\usepackage[version = 4]{mhchem} 

\usepackage{hyperref}
\usepackage{placeins}
\usepackage{authblk}
\author[1,2]{Vivian J. Santamaria-Garcia}
\author[3]{Morgan G. Blevins}
\author[1]{Simo Pajovic}
\author[1,4]{Carolina Orona-Navar}
\author[1]{Svetlana V. Boriskina$^\ast$}
\affil[1]{Department of Mechanical Engineering, Massachusetts Institute of Technology, Cambridge, MA 02139, United States}
\affil[2]{Tecnologico de Monterrey, Escuela de Ingeniería y Ciencias, Monterrey, NL 64700, Mexico}
\affil[3]{Department of Electrical Engineering and Computer Science, Massachusetts Institute of Technology, Cambridge, MA 02139, United States}
\affil[4]{Tecnologico de Monterrey, Artificial Intelligence for Manufacturing and Supply Chain Institute, Monterrey, NL 64700, Mexico}

\def\scititle{Visible-NIR-Frequency Hyperbolic Response in Nodal-Line Semimetal PbTaSe$_2$}
\title{\scititle}
\date{*Email: sborisk@mit.edu}

\begin{document}
\begin{refsection}
\maketitle

\begin{abstract}
\noindent Natural hyperbolic materials offer a powerful platform for light–matter interactions by supporting highly anisotropic electromagnetic modes without the need for artificial patterning. In this work, we experimentally demonstrate that the nodal-line semimetal PbTaSe{$_2$} exhibits robust hyperbolic optical behavior in the visible to near-infrared spectral range, which arises intrinsically from its spin-orbit-coupled electronic structure and layered crystal symmetry. By combining first-principles calculations, ellipsometry, Drude-Lorentz modeling, and reflectance measurements, we establish a consistent and experimentally validated picture of bulk hyperbolicity in this material. This hyperbolicity is of plasmonic origin and is characterized by a competitive quality factor ($Q_\mathrm{max}\approx 2.8$) and a very large anisotropy parameter ($|R|\approx231$) at $0.78$~eV.
\end{abstract}

\section*{Keywords}

Natural hyperbolic materials; nodal-line semimetals; imaging spectroscopic ellipsometry; optical anisotropy.

\section*{Abbreviations}

AFM, atomic force microscopy;  
DFT, density functional theory;  
DL, Drude--Lorentz;  
FTIR, Fourier-transform infrared spectroscopy;  
ISE, imaging spectroscopic ellipsometry;  
NIR, near-infrared;  
PDOS, projected density of states;  
SI, Supporting Information;  
SOC, spin--orbit coupling.  

\section{Introduction}

Extreme confinement and directional control of light at the nanoscale remain central challenges in modern photonics, with applications ranging from on-chip optical interconnects to biosensing and quantum information processing. Addressing these challenges requires material platforms capable of supporting and manipulating electromagnetic fields beyond the diffraction limit. Hyperbolic materials, in which the diagonal components of the dielectric tensor have opposite signs along different axes, exhibit open hyperboloid isofrequency surfaces that enable high-momentum propagation, negative refraction, and sub-diffraction focusing of light~\cite{Narimanov2015,Tamkur_intro}. While artificial metamaterials were the first systems to demonstrate such optical responses, the discovery of natural hyperbolic materials has opened new opportunities for broadband, low-loss, and tunable photonic platforms~\cite{Hao2025}.

Topological semimetals have emerged as a promising platform for highly confined and directional light propagation. These quantum materials host symmetry-protected band crossings in their electronic band structure near the Fermi level. When combined with strong spin-orbit coupling and crystal anisotropy, these topological features generate anisotropic electronic responses~\cite{PhysRevB.107.035114}. This electronic topology is predicted to manifest itself directly in the optical conductivity tensor, resulting in a hyperbolic dispersion that emerges intrinsically from the electronic band structure, in contrast to photonic metamaterials where hyperbolicity is engineered through geometric patterning~\cite{PhysRevLett.131.096902, Passler2022}. Several transition-metal dichalcogenides and nodal-line semimetals, including WTe$_2$~\cite{Wang2020} MoTe$_2$~\cite{Edalati-Boostan_Cocchi_Draxl_2020}, NiTe$_2$~\cite{Rizza_2022}, and ZrSiSe~\cite{Shao_2022} have shown directional permittivity sign changes arising from their anisotropic electronic band structures influenced by topology. 

Despite the rich physics and high potential of topological semimetals for nanophotonic applications, establishing their hyperbolic optical response across experimentally accessible frequency ranges remains an important challenge. A majority of reported hyperbolic systems rely either on phonon polaritons in polar dielectrics such as h-BN ~\cite{Dai2015} and MoO$_3$ ~\cite{doi:10.1126/sciadv.aav8690} or on surface polariton modes in engineered metal-dielectric composites ~\cite{Poddubny2013}. Intrinsic electronic hyperbolicity in layered semimetals requires materials with the right combination of anisotropy, spin-orbit-coupling, and favorable loss characteristics, particularly in the visible spectrum, which is interesting for many photonic applications~\cite{Wang2024}.

PbTaSe$_2$ is a non-centrosymmetric layered topological semimetal that hosts a symmetry-protected nodal line in its electronic band structure arising from strong spin-orbit coupling and broken inversion symmetry. Previous studies have focused on its superconductivity and topological band crossings~\cite{Bian2015TopologicalNF,PhysRevB.93.054520}, yet its anisotropic optical properties remain largely unexplored. Given its layered structure and metallic character, PbTaSe$_2$ is a natural candidate to exhibit hyperbolic dispersion and support plasmon-polariton modes. 

Here we combine broadband spectroscopic ellipsometry and first-principles calculations to determine the full dielectric tensor of PbTaSe$_2$. Anisotropic Drude--Lorentz modeling of the experimental ellipsometric response reveals spectral regions where the in-plane and out-of-plane components of the real permittivity exhibit opposite signs. Experimentally, Type-II hyperbolic regions are identified between $\approx 0.8-1.4$~eV and $1.9-2.6$~eV, spanning the near-infrared and visible spectral ranges. Corresponding \textit{ab initio} calculations predict Type-II regions between $0.8-1.1$~eV and $4.7-5.3$~eV. The experimentally extracted imaginary components ($\mathrm{Im}[\varepsilon] \approx 2-15$) yield a maximum hyperbolic quality factor of $Q_{\mathrm{max}} \approx 2.8$, while first-principles calculations predict a lower value of $Q_{\mathrm{max}} \approx 0.34$. These values indicate that PbTaSe$_2$ maintains appreciable optical confinement despite dissipative losses and can support hyperbolic plasmon-polariton modes. These findings establish PbTaSe$_2$ as a natural hyperbolic topological material operating across the near-infrared and visible spectral ranges. Comparative analysis with other natural hyperbolic materials further shows that PbTaSe$_2$ combines experimentally accessible Type-II hyperbolicity at relatively low photon energies with competitive hyperbolic quality factors and strong optical anisotropy. Together, these results highlight nodal-line semimetals as a promising platform for hyperbolic nanophotonics and plasmonic light manipulation without the need for geometric metamaterial structuring.

\section{Results and discussion}

\subsection*{First principles calculations and relationship to optical response}

PbTaSe$_2$ crystallizes in a non-centrosymmetric hexagonal structure (space group $P\bar{6}m2$, No.~187, $D{3h}$), as shown in Fig.~\ref{fig:Crystal-Bands}A. The layered stacking of TaSe$_2$ and Pb units defines inequivalent in-plane (ab-plane) and out-of-plane (c-axis) crystallographic directions. This structural motif introduces a broken inversion symmetry while maintaining strong in-plane covalency, which plays a central role in its topological and optical anisotropy. 

Prior X-ray diffraction experiments established the lattice parameters of PbTaSe$_2$ as $a=b=3.45$~\AA\ and $c=9.35$~\AA\ with Pb (1a), Ta (1d), and Se (2g) Wyckoff positions\cite{Exp_Lattice_KumarVivek2025}. Our modeling using density functional theory (DFT) yields a relaxed structure, which preserves this symmetry and stacking sequence, yielding $a=b=3.43$~\AA\ and $c=9.37$~\AA, in good agreement with experiment. The crystal symmetry constrains the second-rank dielectric tensor to the uniaxial form

\begin{equation} \label{eq:uniaxial}
    \overline{\overline{\varepsilon}}=
    \begin{bmatrix}
    \varepsilon_{ab} & 0 & 0 \\
    0 & \varepsilon_{ab} & 0 \\
    0 & 0 & \varepsilon_{c}
    \end{bmatrix},
\end{equation}
where $\varepsilon_{ab}$ and $\varepsilon_c$ represent the in-plane and out-of-plane dielectric components, respectively. This symmetry-imposed anisotropy establishes the framework for interpreting the optical response.

Spin-orbit coupling (SOC) plays a critical role in shaping the electronic states that participate in transport and optical transitions near the Fermi level (Fig.~\ref{fig:Crystal-Bands}B). In many nodal-line semimetals, SOC is known to gap nodal-line states, particularly in systems where these crossings are protected only by inversion ($P$) and time-reversal ($T$) symmetries, with the gap magnitude increasing with the average atomic weight of the constituent elements~\cite{Zhang2019}. In contrast, PbTaSe$_2$ is a notable exception: despite its heavy elements, the nodal-line features remain robust in the presence of SOC due to additional crystalline symmetry protection in its non-centrosymmetric $P\bar{6}m2$ structure~\cite{Bian2015TopologicalNF}. Rather than destroying the nodal-line-derived states, SOC lifts degeneracies and reconstructs the Fermi surface, leading to reshaped pockets and modified energy-momentum distributions near the Fermi level. This reconstruction alters the polarization dependence of optical transitions, directly impacting the observed anisotropic optical response (Fig. \ref{fig:SOC-effect}).

Figure~\ref{fig:Crystal-Bands}C provides direct insight into the microscopic origin of optical anisotropy through orbital-projected band structures and the projected density of states (PDOS). States near the Fermi level are dominated by Ta $d_{xy}$ and $d_{x^2-y^2}$ orbitals, with contributions from Pb and Se $p_{x}$ and $p_{y}$ states. These orbitals couple efficiently to in-plane electric fields, resulting in a strong metallic response within the basal plane. In contrast, the comparatively weaker contribution of orbitals with out-of-plane character suppresses optical transitions polarized along the c-axis. Figure~\ref{fig:Crystal-Bands}D further clarifies the role of SOC by zooming in on representative band crossings along high-symmetry directions. SOC lifts degeneracies and shifts interband transition energies differently along in-plane and out-of-plane momentum directions (Fig. \ref{fig:BZ}). These SOC-induced splittings redistribute optical oscillator strength and set the energy scales at which distinct dielectric tensor components evolve and may undergo sign changes.

\subsection*{Ellipsometric characterization}

\begin{figure}[H]
	\centering
	\includegraphics[width=\textwidth]{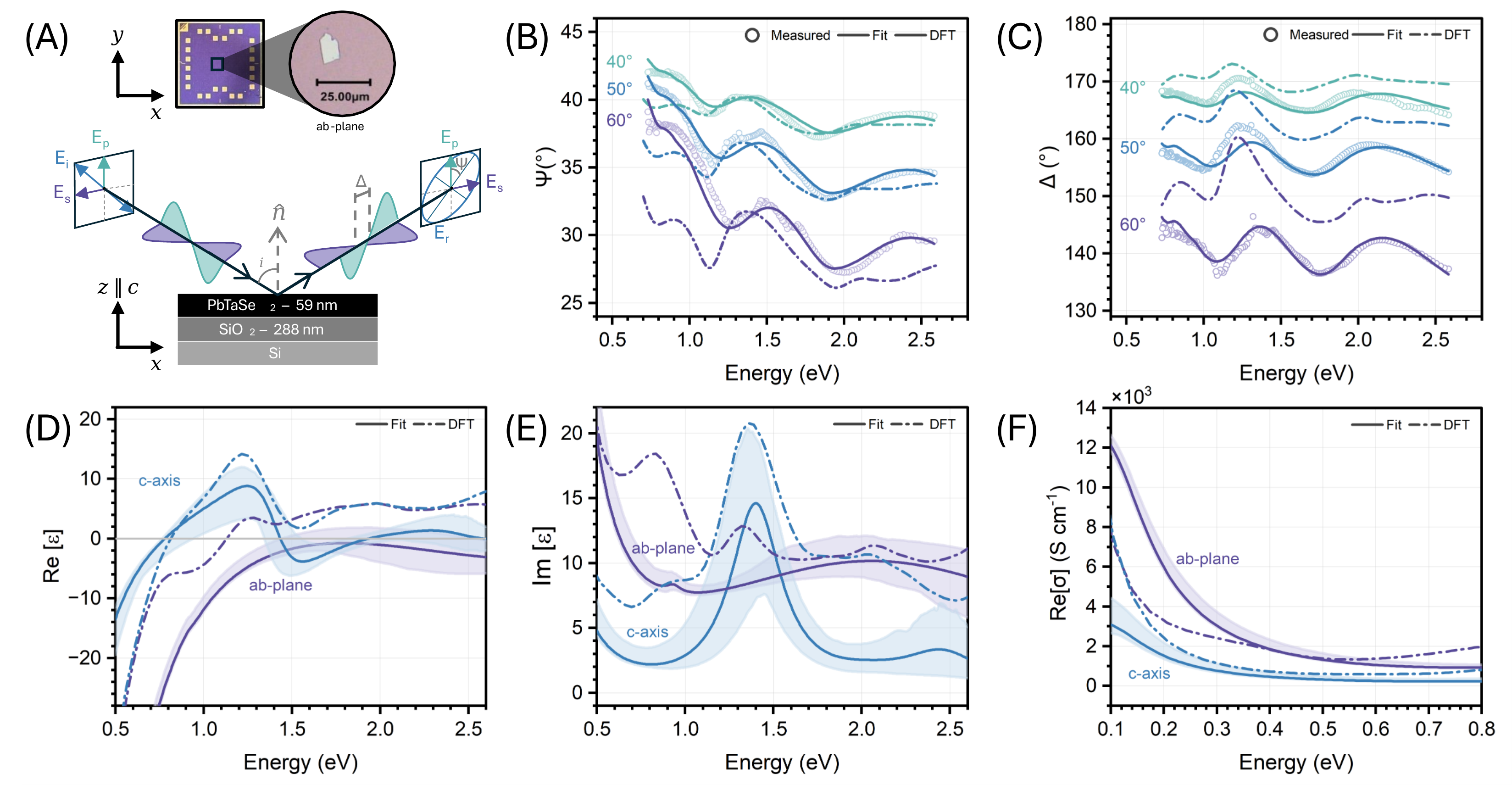}

    \caption{Uniaxial anisotropic dielectric response of PbTaSe$_2$ (A) A micrograph of the transferred flake and a schematic of the multi-layer structure used for fitting ellipsometric data. (B) Ellipsometric angle  $\Psi$ measured at angles of incidence of 40°, 50°, and 60°, fitted using an anisotropic DL model consisting of two Lorentz oscillators and one Drude term for $\varepsilon_{ab}$ and three Lorentz oscillators and one Drude term for $\varepsilon_{c}$ (Table \ref{tab:anisotropic_DL_model}). (C) Corresponding spectra of ellipsometric angle $\Delta$ at the same angles of incidence, showing good agreement with the optical model. (D,E) Real and imaginary parts of the in-plane dielectric function, $\varepsilon_{ab}$, extracted from the experimental data using a uniaxial anisotropic DL model, showing a strong metallic response dominated by free carriers. The experimentally extracted anisotropic dielectric functions and first-principles calculations show good agreement and confirm the intrinsic origin of the optical anisotropy. (F) Real part of the optical conductivity, $\mathrm{Re}[\sigma]$, for the in-plane ($ab$) and out-of-plane ($c$) directions, derived from the fitted dielectric functions. The in-plane response is dominated by a strong low-energy Drude contribution, indicating free-carrier transport, while the out-of-plane conductivity remains significantly reduced, reflecting suppressed interlayer transport and anisotropic carrier dynamics. Shaded regions around the fit represent the 95 \% confidence intervals (CI)).}
	\label{fig:Anisotropic-fit}
\end{figure}

The potential for a natural, highly anisotropic optical response originating from the dominant contributions of in-plane orbitals motivates ellipsometric characterization. Thin flakes of PbTaSe$_2$ were mechanically exfoliated from a bulk crystal and transferred to an silica-on-silicon (SiO{$_2$}-on-Si) wafer. Spectroscopic ellipsometry measurements were performed at angles of incidence of 40$^\circ$, 50$^\circ$ and 60$^\circ$, measuring the ellipsometric parameters $\Psi$ and $\Delta$ as a function of photon energy (Fig.~\ref{fig:Anisotropic-fit}A). No Mueller matrix analysis was conducted. The optical response was modeled using the uniaxial dielectric tensor of Eq.~\eqref{eq:uniaxial}, assuming the crystallographic c-axis to be oriented normal to the sample surface. This assumption is justified by the layered van der Waals structure of PbTaSe$_2$, where mechanical exfoliation occurs along well-defined cleavage planes, preserving the c-axis orientation normal to the surface while maintaining in-plane symmetry.

The measured structure was modeled as a multilayer system consisting of a Si substrate, a 287~nm-thick intermediate SiO$_2$ layer, and a 59~nm-thick top PbTaSe$_2$ layer. The SiO$_2$ thickness is fixed to the nominal 287~nm value (consistent with standard thermally grown oxide, Supporting Fig.~\ref{fig:SiO2-thickness}), while the PbTaSe$_2$ thickness is fixed to the experimentally-determined flake thickness ($\sim$59~nm, determined via atomic force microscopy (AFM) as shown in Supporting Fig.~\ref{fig:AFM-PbTaSe2-thickness}). AFM measurements further confirm that the flake is laterally uniform and exhibits negligible tilt ($\theta \approx 0^\circ$) over the area probed by ellipsometry, supporting the treatment of the measured region as a single effective anisotropic domain. Within this multilayer optical model, $\Psi$ and $\Delta$ were simultaneously fit at all three angles of incidence, ensuring that the extracted dielectric functions are consistent across measurement geometries and that interference effects from the oxide layer are properly accounted for.

We then extracted the independent in-plane, $\varepsilon_{ab}$, and out-of-plane, $\varepsilon_{c}$, dielectric components of PbTaSe$_2$ by fitting the $\Psi$ and $\Delta$ datasets using a Drude-Lorentz (DL) model:  
\begin{equation} \label{eq:}
\begin{aligned}
    \varepsilon(E) &= \varepsilon_\infty + \varepsilon_{Drude} + \sum_{j=1}^{l}\varepsilon_{Lorentz,j} \\
    &= \varepsilon_\infty -\frac{A}{E^2+i{\Gamma}E} + \sum_{j=1}^{l} \frac{A_j}{E_{0,j}^2-E^2-i{\Gamma_j}E}. 
\end{aligned}
\end{equation}
The results of the fit are in Table~\ref{tab:anisotropic_DL_model} and plotted in Fig.~\ref{fig:Anisotropic-fit}D-E.

\begin{table}
\centering
\caption{\textbf{Anisotropic Drude-Lorentz model parameters.}\\ab-plane: $\varepsilon_\infty=0.500$; c-axis: $\varepsilon_\infty=3.369$.}
\vspace{0.15cm}
\begin{tabular}{c|ccc||ccc}
\hline
 & \multicolumn{3}{c||}{\textbf{ab-plane}, $\varepsilon_{ab}$} & \multicolumn{3}{c}{\textbf{c-axis}, $\varepsilon_{c}$} \\
Oscillator & E$_0$ (eV) & A (eV$^2$) & $\Gamma$ (eV) & E$_0$ (eV) & A (eV$^2$) & $\Gamma$ (eV) \\
\hline
L1 & 0.935 & 0.059  & 0.104 & 1.410 & 6.944 & 0.350 \\
L2 & 2.542 & 70.013  & 3.050 & 2.051 & 0.745 & 0.559 \\
L3 & -- & -- & -- & 2.475 & 3.890 & 0.573 \\
D1 & --    & 22.089 & 0.100 & --    & 5.628 & 0.100 \\
\hline
\end{tabular}
\label{tab:anisotropic_DL_model}
\end{table}

Further details on the calculation of the reflection coefficients used in the optical model are provided in the Supporting Information. The $\varepsilon_{ab}$ and $\varepsilon_{c}$ components were modeled using independent oscillator sets and fitted simultaneously within a single global optimization. This fit captures the directional dependence expected from the layered structure and anisotropic electronic dispersion. Although Mueller-matrix ellipsometry would provide access to depolarization effects and deviations from ideal uniaxial behavior, the combination of (i) multiple angles of incidence, (ii) a well-defined multilayer stack (Si/SiO$_2$/PbTaSe$_2$), and (iii) the symmetry-imposed uniaxial tensor form yields a physically consistent extraction of $\varepsilon_{ab}$ and $\varepsilon_c$ over the measured energy window \cite{Choi2026,Ermolaev2021}.

Panels~\ref{fig:Anisotropic-fit}D–F provide a unified view of the anisotropic optical response by combining the measured dielectric functions, first-principles calculations, and derived optical conductivities $\sigma$. The experimental dielectric functions $\varepsilon_{ab}$ and $\varepsilon_c$ show good overall agreement with DFT, capturing the strongly metallic in-plane response and the suppressed, energy-dependent out-of-plane behavior. In particular, the enhanced Drude contribution in $\varepsilon_{ab}$ reflects the dominance of in-plane carrier dynamics, while the reduced Drude weight and delayed interband onset in $\varepsilon_c$ indicate limited interlayer transport and anisotropic optical matrix elements.

The corresponding optical conductivity spectra (Fig.~\ref{fig:Anisotropic-fit}F) provide a complementary transport-oriented perspective. The in-plane component, $\mathrm{Re}[\sigma_{ab}]$, is dominated by a strong low-energy Drude response that rapidly decreases with photon energy, consistent with intraband carrier dynamics near the Fermi level, while $\mathrm{Re}[\sigma_{c}]$ remains significantly smaller, reflecting suppressed out-of-plane transport. This anisotropy mirrors the orbital-resolved electronic structure, where Ta $d$ and in-plane $p$ orbitals dominate near the Fermi level. Such behavior is consistent with optical studies of nodal-line semimetals, where low-energy conductivity is governed by free carriers and higher-energy spectral weight arises from interband transitions across linearly dispersing bands~\cite{Schilling2017,Shao2020}. Spin-orbit coupling further redistributes spectral weight without eliminating the underlying anisotropic metallic response, consistent with the behavior observed in PbTaSe$_2$~\cite{Bian2016}.

While the overall trends are consistent, there is a notable difference between experiment and theory at higher photon energies. The experimental DL fit for $\mathrm{Re}[\varepsilon_{ab}]$ remains negative over the measured spectral range, whereas the first-principles dielectric function becomes positive. This discrepancy arises because the DL model emphasizes the low-energy free-carrier response constrained by the ellipsometric window, while first-principles calculations capture higher-energy interband transitions that progressively dominate the optical response. Additionally, dissipative losses in the experimental system likely contribute to the broadening of these interband features.

Finally, a reflection spectrum measured at normal incidence (635 nm) is in good agreement with both the experimentally extracted and DFT-predicted permittivity spectra (Supporting Fig.~\ref{fig:635nm-reflection}), providing an independent validation of the optical model.

\subsection*{Hyperbolic response}

Figure~\ref{fig:Hyperbolic-ranges}A explicitly identifies the hyperbolic regimes of PbTaSe$_2$ measured experimentally. At low photon energies, the in-plane component $\mathrm{Re}[\varepsilon_{ab}]$ remains strongly negative, reflecting a dominant free-carrier response associated with delocalized conduction within the basal plane. In contrast, the out-of-plane component $\mathrm{Re}[\varepsilon_{c}]$ exhibits a weaker metallic contribution and approaches zero more rapidly with increasing photon energy, consistent with suppressed interlayer transport. This pronounced anisotropy gives rise to a finite spectral window where $\mathrm{Re}[\varepsilon_{ab}] < 0$ and $\mathrm{Re}[\varepsilon_{c}] > 0$, corresponding to Type-II hyperbolic dispersion. No Type-I hyperbolic regime is observed within the experimental energy range.

At higher photon energies, the DL model predicts a second sign change in $\mathrm{Re}[\varepsilon_{c}]$, while $\mathrm{Re}[\varepsilon_{ab}]$ remains negative. Because this feature occurs near the upper limit of the measured range, its physical origin is less certain and may reflect the combined effects of limited spectral coverage and overlapping interband transitions.

Although the exact energies of the sign inversions differ, the first-principles results (Fig.~\ref{fig:Hyperbolic-ranges}B) reproduce the essential condition for hyperbolic dispersion—opposite signs of $\varepsilon_{ab}$ and $\varepsilon_c$. The remaining discrepancies, including the apparent suppression of some interband features, likely arise from the finite experimental spectral window and the overlap of multiple interband contributions within the fitted optical response, which redistribute spectral weight and modify the effective oscillator strengths. Similar discrepancies between experimentally extracted and first-principles hyperbolic responses have also been reported in materials such as Bi$_2$Se$_3$, where the predicted hyperbolic windows are sensitive to carrier concentration, band dispersion, and exchange--correlation approximations~\cite{Bi2Se3_Bi2Te3_Esslinger,Bi2Se3_Exp,Bi2Se3_exp_Lingstdt2021,Bi2Se3_Koc2014}. In this context, the differences between the experimentally extracted and theoretically predicted hyperbolic windows of PbTaSe$_2$ are not unexpected, but rather reflect the sensitivity of electronic hyperbolicity to subtle changes in band structure and carrier dynamics, particularly in systems hosting spin--orbit-modified nodal-line states. 

Overall, the qualitative agreement between experiment and theory supports the identification of the experimentally observed Type-II hyperbolic regime as an intrinsic property of PbTaSe$_2$.

\subsection{Benchmarking Hyperbolic Performance}

From a physical perspective, the emergence of hyperbolicity in PbTaSe$_2$ arises from the coexistence of a strong in-plane metallic response and a comparatively suppressed, energy-dependent out-of-plane response. The layered crystal structure promotes delocalized charge transport within the basal plane, while reduced interlayer coupling limits charge transport and optical absorption along the crystallographic c-axis. 

In PbTaSe$_2$, this intrinsic anisotropy is further influenced by spin-orbit-modified nodal-line bands near the Fermi level. First-principles calculations show that these states are dominated by in-plane Ta $d$ and chalcogen $p$ orbitals, which can interact strongly with in-plane electric fields, consistent with the enhanced $\varepsilon_{ab}$ and $\sigma_{ab}$ observed experimentally, while out-of-plane optical responses remain comparatively suppressed. 

As photon energy increases, additional interband transitions become accessible, with polarization-dependent coupling to in-plane and out-of-plane electric fields. This polarization-selective response drives the distinct evolution of $\varepsilon_{ab}$ and $\varepsilon_{c}$, leading to the observed dielectric sign inversions.

\begin{figure}[H]
	\centering
	\includegraphics[width=\textwidth]{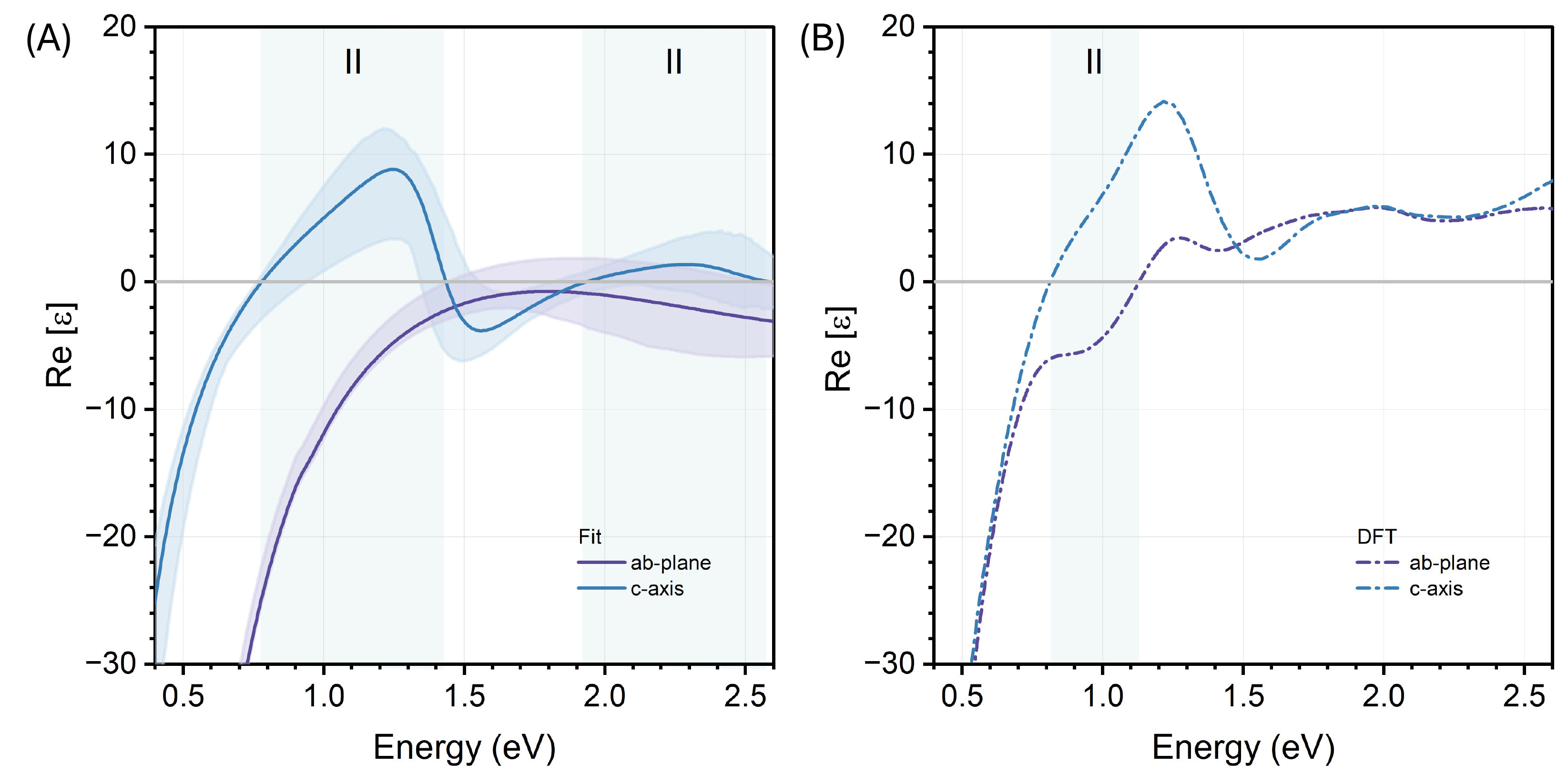}
    \caption{Identification of hyperbolic optical regimes in PbTaSe$_2$. (A) Real parts of the in-plane and out-of-plane dielectric functions from anisotropic DL modeling of experimental data. The shaded regions around the fit represent the 95 \% confidence intervals (CI) (B) Corresponding real parts from first-principles calculations. The vertical shaded regions indicate spectral regions with opposite signs of $\varepsilon_{ab}$ and $\varepsilon_{c}$, where PbTaSe$_2$ has Type-II hyperbolic dispersion.}
	\label{fig:Hyperbolic-ranges}
\end{figure}

The anisotropic DL model reveals finite spectral windows in which $\mathrm{Re}[\varepsilon_{ab}] < 0$ while $\mathrm{Re}[\varepsilon_c] > 0$ (Fig.~\ref{fig:Hyperbolic-ranges}), defining Type-II hyperbolic dispersion within the uniaxial symmetry of PbTaSe$_2$. First-principles calculations also predict Type-II hyperbolic regions, reproducing the experimentally observed near-infrared behavior while differing in the higher-energy spectral range. In this regime, the dielectric sign contrast produces an open hyperboloid isofrequency surface that could enable access to high-momentum electromagnetic modes, enhanced spontaneous emission, and strongly confined optical fields, even in the presence of moderate optical losses~\cite{Jacob06,TopHM,Poddubny2013,Caldwell_lowloss}. Because this hyperbolicity originates from intrinsic electronic anisotropy and extends into the visible--near-infrared range, PbTaSe$_2$ represents a natural platform for electronic hyperbolic nanophotonics without the need for metamaterial structuring.

\begin{figure}[H]
	\centering
	\includegraphics[width=\textwidth]{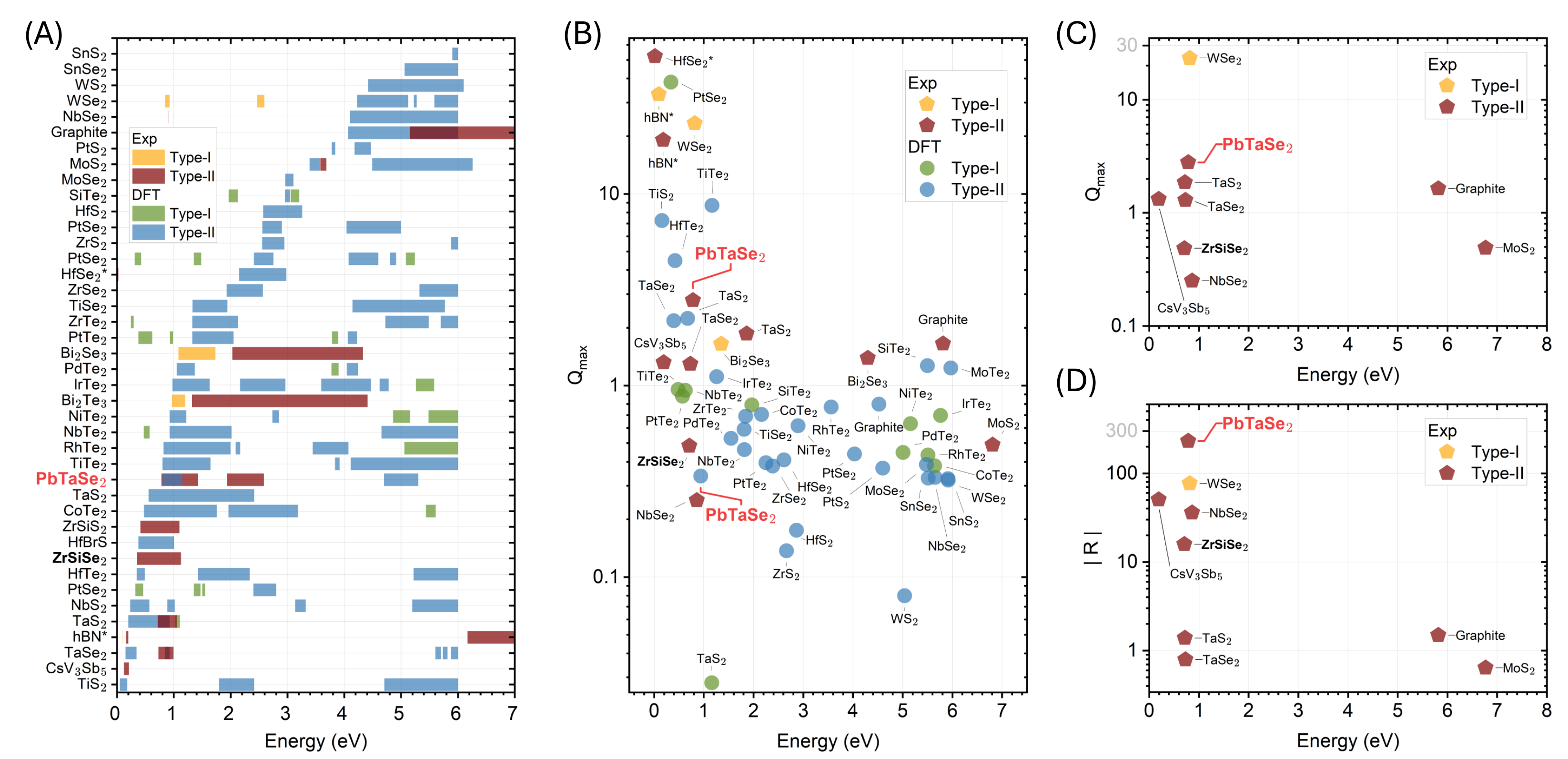}
    \caption{Spectral regions and performance of natural uniaxial hyperbolic van der Waals materials. (A) Type-I and Type-II hyperbolic spectral regions as predicted by first-principles calculations (green and blue) and as measured experimentally (yellow and red). (B) Maximum hyperbolic quality factor $\textit{Q}$ as predicted by first-principles calculations vs. experimental measurements (Table ~\ref{tab:natural-hyperbolic-materials})~\cite{GNat_Comm_DFT, PtSe2_Ghasemi2020, Bi2Se3_Bi2Te3_Esslinger,hBN_Caldwell2014,graphite_exp,NbSe2_TaSe2_exp,WSe2_exp,HfSe2_exp,MoS2_exp,Shao_2022,Shiravi2024}. (C) Maximum hyperbolic quality factor $Q_{\max}$ as a function of energy considering only experimental reports where the hyperbolic response is attributed to or strongly associated with plasmonic behavior, emphasizing the trade-off between operating energy and optical performance. (D) Ratio of the real parts of the dielectric function, $|R| = |\mathrm{Re}[\varepsilon_{ab}]/\mathrm{Re}[\varepsilon_c]|$, evaluated at the energy corresponding to the maximum $Q$, highlighting the degree of optical anisotropy associated with peak hyperbolic performance. Labels marked with ($\ast$) in A and B indicate materials for which a phonon-driven origin of hyperbolicity has been explicitly reported in the literature; this designation does not exclude other materials from exhibiting phonon-related contributions. Bold labels denote topological nodal-line semimetals.}
	\label{fig:Comparative}
\end{figure}

Figure~\ref{fig:Comparative} places PbTaSe$_2$ in the broader context of natural uniaxial van der Waals hyperbolic materials, including hBN, graphite, and representative transition-metal dichalcogenides~\cite{GNat_Comm_DFT, PtSe2_Ghasemi2020, Bi2Se3_Bi2Te3_Esslinger, hBN_Caldwell2014,graphite_exp, NbSe2_TaSe2_exp,WSe2_exp,HfSe2_exp,MoS2_exp}. These materials share a common layered structural motif characterized by strong in-plane bonding and weak interlayer van der Waals interactions, which give rise to distinct in-plane and out-of-plane optical responses and enable natural hyperbolic behavior without geometric patterning. Panel ~\ref{fig:Comparative}A summarizes hyperbolic behavior only within the spectral ranges reported in the cited literature. The indicated hyperbolic windows therefore reflect the available dielectric data and do not exclude the possibility of additional hyperbolic regions outside the investigated energy intervals. It is remarkable that most materials exhibit hyperbolicity either in narrow phonon-driven mid-IR bands (e.g., hBN ~\cite{hBN_Caldwell2014}) or at higher energies where losses are significant. In contrast, PbTaSe$_2$ displays experimentally confirmed Type-II hyperbolicity in the visible–NIR region.

To quantify the trade-off between optical confinement and dissipation, we evaluate the quality factor $Q$, following the established framework for hyperbolic media~\cite{Korzeb_Optika}. It is calculated for the direction in which the material exhibits metallic behavior ($\varepsilon_{ab}$ in the case of PbTaSe$_2$) as:

\begin{equation}
Q = \frac{|\mathrm{Re}[\varepsilon]|}{\mathrm{Im}[\varepsilon]},
\end{equation}

\noindent where $\mathrm{Re}[\varepsilon]$ governs phase accumulation and wavevector enhancement, while $\mathrm{Im}(\varepsilon)$ captures absorptive damping. A large $Q$ indicates strong confinement relative to losses, whereas a small $Q$ corresponds to overdamped excitations. The maximum value, $Q_{\mathrm{max}}$, therefore identifies the spectral regions where the material most efficiently supports confined hyperbolic modes. For PbTaSe$_2$, the experimentally extracted response yields $Q_{\mathrm{max}} \approx 2.8$ at $\approx 0.78$~eV within the visible--NIR hyperbolic window, consistent with the Drude-dominated in-plane response and moderate optical losses. In comparison, first-principles calculations predict a lower value of $Q_{\mathrm{max}} \approx 0.34$ at $\approx 0.94$~eV.

Panel ~\ref{fig:Comparative}B compares the $Q_{\mathrm{max}}$, calculated from the experimental and DFT dielectric data available for the materials summarized in panel~\ref{fig:Comparative}A. While phonon-polaritonic systems can achieve larger $Q$ values at mid-IR frequencies, the experimentally extracted response of PbTaSe$_2$ maintains competitive $Q_{\mathrm{max}}$ values at visible energies, where electronic damping typically limits performance in electronic hyperbolic materials. In contrast, the DFT-derived $Q_{\mathrm{max}}$ for PbTaSe$_2$ is substantially smaller, reflecting the stronger interband contributions and reduced effective confinement predicted by the first-principles dielectric response.

Panels~\ref{fig:Comparative}C and \ref{fig:Comparative}D further clarify this trade-off using only experimentally reported hyperbolic materials with predominantly electronic or plasmon-polaritonic character. Panel~\ref{fig:Comparative}C compares $Q_{\mathrm{max}}$ for these materials, showing that PbTaSe$_2$ exhibits a competitive hyperbolic quality factor at relatively low photon energy. Panel~\ref{fig:Comparative}D compares the anisotropy parameter, defined here as $|R| = |\mathrm{Re}[\varepsilon_{ab}]/\mathrm{Re}[\varepsilon_c]|$ and evaluated at the energy corresponding to $Q_{\mathrm{max}}$. This comparison shows that PbTaSe$_2$ combines moderate $Q_{\mathrm{max}}$ with strong optical anisotropy, a balance that is particularly notable among electronic hyperbolic systems where achieving both visible-frequency operation and appreciable confinement remains challenging.

Overall, the coexistence of topologically-derived electronic anisotropy, moderate dissipation, and broadband Type-II hyperbolic dispersion suggests that thin PbTaSe$_2$ layers can support hyperbolic plasmon–polaritons with subwavelength confinement in the visible–NIR regime. This identifies PbTaSe$_2$ as a promising hyperbolic material governed by free carriers and spin–orbit-coupled electronic structure, rather than lattice resonances, bridging quantum materials and nanophotonics and offering a pathway toward tunable optoelectronic and plasmonic devices.\\


\begin{figure} [H]
	\centering
	\includegraphics[width=\textwidth]{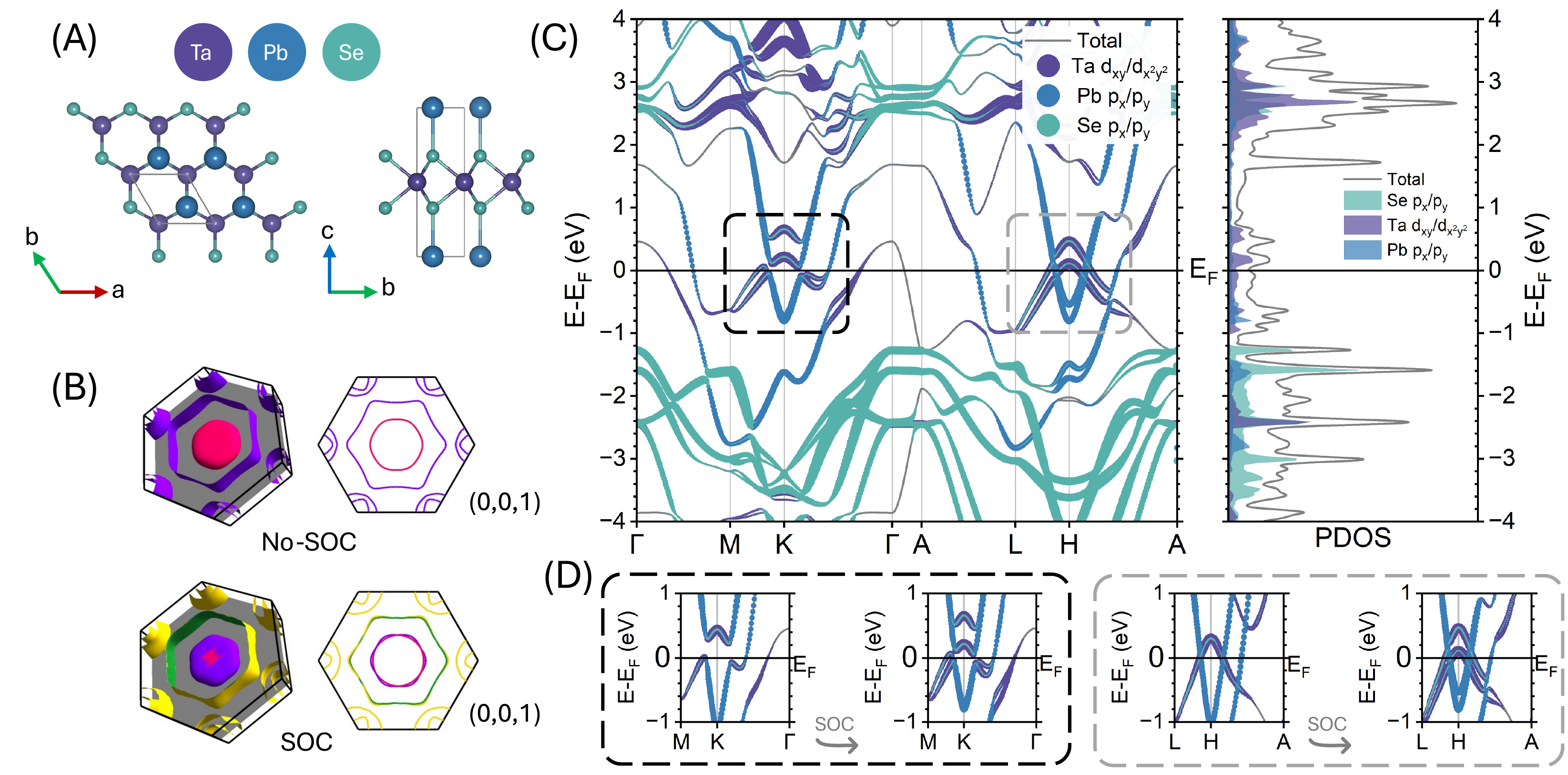}
    \caption{Electronic structure of PbTaSe$_2$ and spin–orbit coupling effects (A) Crystal structure of PbTaSe$_2$, highlighting its layered hexagonal symmetry and the inequivalent in-plane (ab) and out-of-plane (c-axis) directions that enable anisotropic electronic dispersion and optical response. (B) Fermi surface calculated without (top) and with (bottom) SOC, showing substantial SOC-induced reconstruction of low-energy electronic states near the Fermi level, including lifted degeneracies and reshaped pockets associated with nodal-line topology. (C) Orbital-projected band structure and density of states, indicating dominant Ta d-orbital character near the Fermi level. (D) SOC-induced band splitting along high-symmetry directions.} 
	\label{fig:Crystal-Bands}
\end{figure}

\section*{Methods}
\subsection*{Sample preparation}
Single crystals of PbTaSe$_2$ were obtained from 2D Semiconductors Inc. Thin flakes were prepared by mechanical exfoliation of bulk PbTaSe$_2$ crystals using adhesive tape and subsequently transferred onto Si wafers coated with 285 nm of thermally grown SiO$_2$. Before transfer, the SiO$_2$/Si substrates were cleaned by using a mild atmospheric plasma treatment to improve adhesion. 

\subsection*{Atomic force microscopy (AFM)}
The thickness and topographic features of the exfoliated PbTaSe$_2$ flakes were characterized using an Asylum Research Jupiter XR atomic force microscope operated in tapping mode. Silicon cantilevers (AC160, Olympus) with a nominal resonance frequency of $\sim$300~kHz and tip radius smaller than 10~nm were used. Height profiles were obtained across the flake edges to determine its thickness, as shown in Fig.~\ref{fig:AFM-PbTaSe2-thickness} in the Supplementary Materials.

\subsection*{Spectroscopic ellipsometry}
Optical characterization was performed by imaging spectroscopic ellipsometry to evaluate the dielectric response of the PbTaSe$_2$ flake and substrate. Measurements were conducted using an Accurion EP4 spectroscopic ellipsometer in the spectral range of 480 to 900 nm with a step of 1 nm and from 900 to 1700 with a step of 5 nm. The reflected beam was collected through a 20× objective included with the imaging ellipsometer. Ellipsometric spectra were acquired at multiple angles of incidence (40$^{\circ}$, 50$^{\circ}$, and 60$^{\circ}$). The ellipsometric parameters $\Psi$ and $\Delta$ were obtained as wavelength functions and fitted using Fresnel equations for multilayer systems to extract the optical constants. The fitting model included the Si substrate, the 288 nm SiO$_2$ layer (Supporting Fig.~\ref{fig:SiO2-thickness}), and the PbTaSe$_2$ flake with its measured thickness from AFM (Supporting Fig.~\ref{fig:AFM-PbTaSe2-thickness}).

\subsection*{FTIR measurements}
Specular reflectance at normal incidence was measured using a Thermo Fisher Nicolet\texttrademark~iS50 FTIR spectrometer equipped with a Continuum FTIR microscope accessory and an MCT-A detector. A gold mirror (Thorlabs PFSQ20-03-M02) was used to measure the reference spectrum before measuring the reflected intensity by the sample using the same aperture size as the reference spectrum. The specular reflectance was calculated as the reflected intensity divided by the reference spectrum.

\subsection*{First Principles Calculations }
First-principles calculations were performed using density functional theory (DFT) within the projector-augmented wave method \cite{PAW} as implemented in the Vienna \textit{Ab Initio} Simulation Package (VASP), employing the Perdew--Burke--Ernzerhof (PBE) generalized gradient approximation \cite{PBE}. A plane-wave energy cutoff of 520~eV was used. Structural relaxations allowed both lattice parameters and atomic positions to relax until residual forces were below \(10^{-3}\,\mathrm{eV}\,\text{\AA}^{-1}\), with van der Waals interactions treated using the DFT-D3 method with Becke--Johnson damping \cite{DFT-D3}. The frequency-dependent dielectric function was computed using the linear optical response formalism on the relaxed structure. The Brillouin zone was sampled with a \(\Gamma\)-centered \(15 \times 15 \times 9\) Monkhorst--Pack \textit{k}-point mesh, and spin--orbit coupling was included.

\section*{Acknowledgements}
The authors thank Dr. Steven E. Kooi for valuable discussions and technical guidance related to ellipsometry and AFM characterization; Dr. Mangesh S. Diware for insightful discussions and troubleshooting on ellipsometry measurements; Abhishek Mukherjee for fabricating and providing dies used in sample preparation; and Christopher Hogan for AFM training and technical support at the MIT--ISN facilities

\paragraph*{Funding:}   This work was supported in part by the MIT MISTI-Poland Seed Fund, and carried out in part through the use of MIT.nano and MIT-ISN facilities. V.J.S. acknowledges support from the MIT–Tecnol\'ogico de Monterrey funding program.  M.G.B. was supported as a Draper Scholar by the Charles Stark Draper Laboratory. Inc. S.P. gratefully acknowledges support from the MathWorks Engineering Fellowship via the Massachusetts Institute of Technology. C.O. acknowledges the generous financial support from Fundaci\'on FEMSA through the \textit{Materials of the Future} project. Additional support from the MIT–Tecnol\'ogico de Monterrey collaboration enabled the research exchange that contributed to this work.

\paragraph*{Author contributions:}

All authors contributed to the conceptualization of the project under the supervision of S.V.B. Sample preparation was carried out by V.J.S., C.O., and M.G.B. Ellipsometry measurements were performed by V.J.S., C.O., and S.P. AFM characterization was conducted by V.J.S. and C.O., while FTIR measurements were carried out by S.P., V.J.S., and C.O. Reflection measurements were performed by M.G.B. First-principles calculations were performed by V.J.S. Optical modeling was conducted by V.J.S., M.G.B., and S.P. All authors contributed to writing, reviewing, and editing the manuscript.

\paragraph*{Competing interests:} The authors have no competing interests to declare.

\paragraph*{Data and materials availability:}
All data needed to evaluate the conclusions in the paper are present in the paper and/or the Supplementary Materials.

\section*{Supporting information}

The following files are available free of charge.
\begin{itemize}
  \item \nameref{sec:brillouin-zone}; \nameref{sec:sio2-thickness}; \nameref{sec:afm}; \nameref{sec:iso-fit}; \nameref{sec:SOC-dielectric-func}; \nameref{sec:refl-measurement}; \nameref{sec:natural-hyperbolic-materials}.
\end{itemize}

\printbibliography
\end{refsection}

\newpage

\begin{refsection}

\setcounter{secnumdepth}{1}
\setcounter{section}{0}

\renewcommand{\thesection}{S\arabic{section}}
\renewcommand{\theHsection}{S\arabic{section}}

\renewcommand{\thefigure}{S\arabic{figure}}
\renewcommand{\thetable}{S\arabic{table}}
\renewcommand{\theequation}{S\arabic{equation}}
\renewcommand{\thepage}{S\arabic{page}}

\setcounter{figure}{0}
\setcounter{table}{0}
\setcounter{equation}{0}

\begin{center}
\section*{Supplementary Materials for\\ \scititle}
Vivian J. Santamaria-Garcia, Morgan G. Blevins, Simo Pajovic, Carolina Orona-Navar, Svetlana V. Boriskina$^{\ast}$
\\
\small$^\ast$Corresponding author. Email: sborisk@mit.edu\\
\end{center}

\vspace{1.5em}
\noindent\rule{\linewidth}{0.4pt}
\vspace{0.3em}
\noindent{\large\textbf{Contents}}
\vspace{0.3em}

\noindent \nameref{sec:brillouin-zone} \dotfill \pageref{sec:brillouin-zone}\\[0.4em]
\noindent \nameref{sec:sio2-thickness} \dotfill \pageref{sec:sio2-thickness}\\[0.4em]
\noindent \nameref{sec:afm} \dotfill \pageref{sec:afm}\\[0.4em]
\noindent \nameref{sec:iso-fit} \dotfill \pageref{sec:iso-fit}\\[0.4em]
\noindent \nameref{sec:SOC-dielectric-func} \dotfill \pageref{sec:SOC-dielectric-func}\\[0.4em]
\noindent \nameref{sec:refl-measurement} \dotfill \pageref{sec:refl-measurement}\\[0.4em]
\noindent \nameref{sec:natural-hyperbolic-materials} \dotfill \pageref{sec:natural-hyperbolic-materials}\\[0.4em]
\noindent\rule{\linewidth}{0.4pt}

\newpage

\section{Brillouin zone and high-symmetry k-path} \label{sec:brillouin-zone}

\begin{figure}[H]
	\centering
	\includegraphics[width=0.5\linewidth]{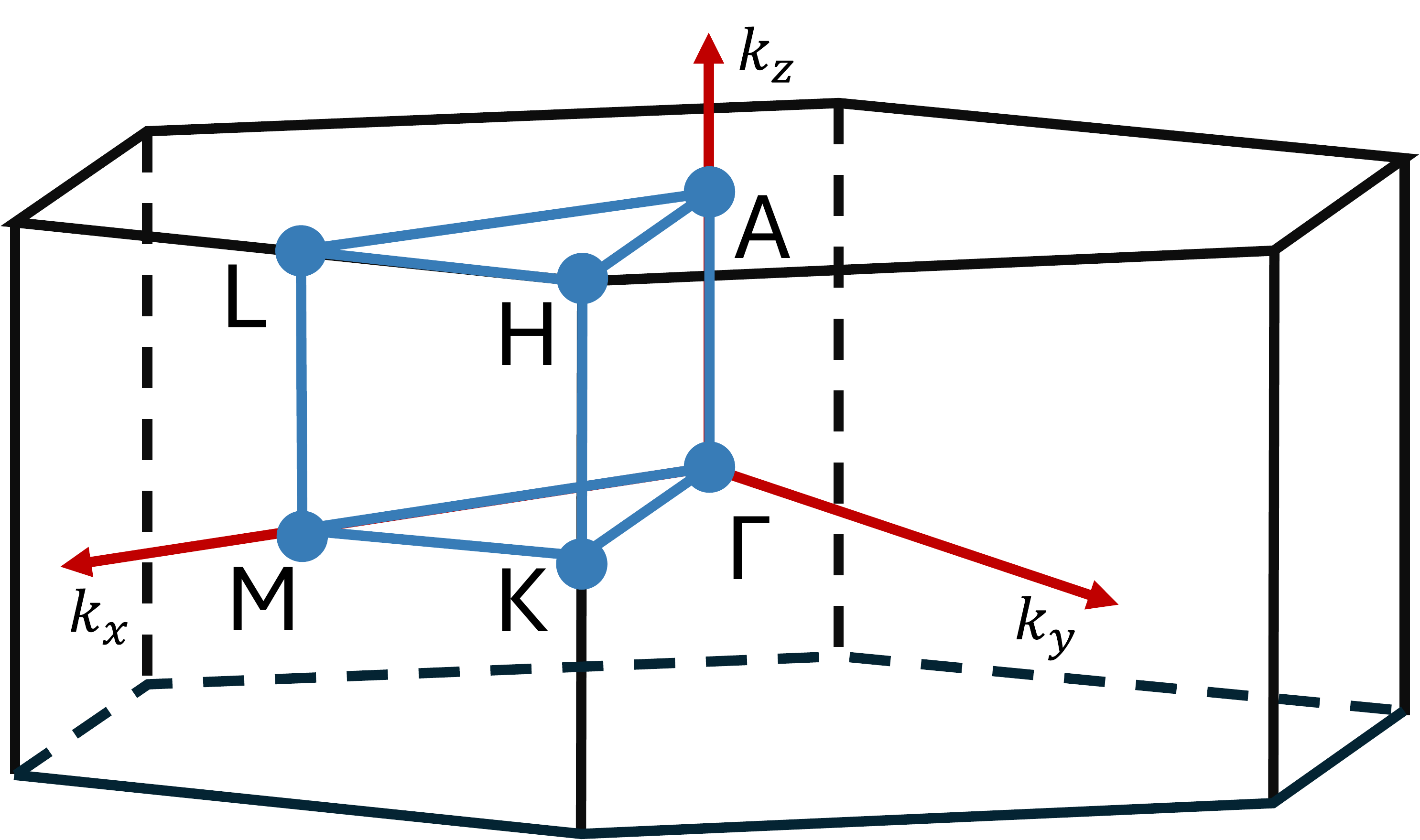}
    \caption{Brillouin zone of hexagonal PbTaSe$_2$ showing the high-symmetry k-path used for the band structure calculations: $\Gamma-M-K-\Gamma-A-L-H-A$.}
	\label{fig:BZ}
\end{figure}

\newpage

\section{Ellipsometric characterization of \texorpdfstring{SiO$_2$}{SiO2} thickness} \label{sec:sio2-thickness}

\begin{figure}[H]
	\centering
	\includegraphics[width=\textwidth]{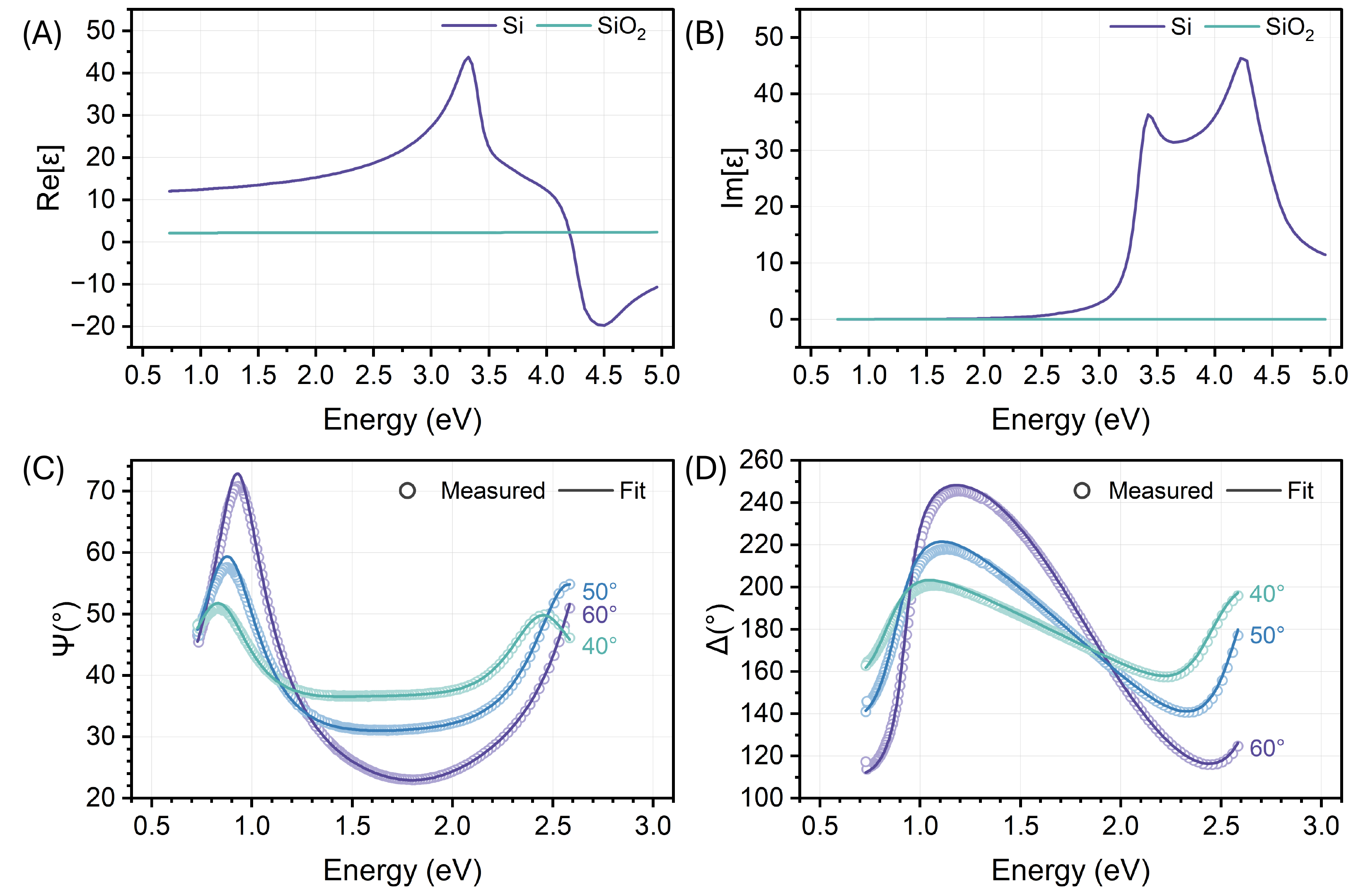}
    \caption{Ellipsometric determination of the SiO$_2$ thickness using a multilayer optical model. (A) Real and (B) imaginary parts of the dielectric functions of Si and SiO$_2$ used as input parameters.(C) Experimental and fitted $\Psi$ and (D) $\Delta$ spectra measured at 40°, 50°, and 60°. The fit yields a SiO$_2$ thickness of 288 nm.}
	\label{fig:SiO2-thickness}
\end{figure}

\newpage

\section{AFM-measured  \texorpdfstring{PbTaSe$_2$}{PbTaSe2} flake thickness} \label{sec:afm}

\begin{figure}[H]
    \centering
    \includegraphics[width=\textwidth]{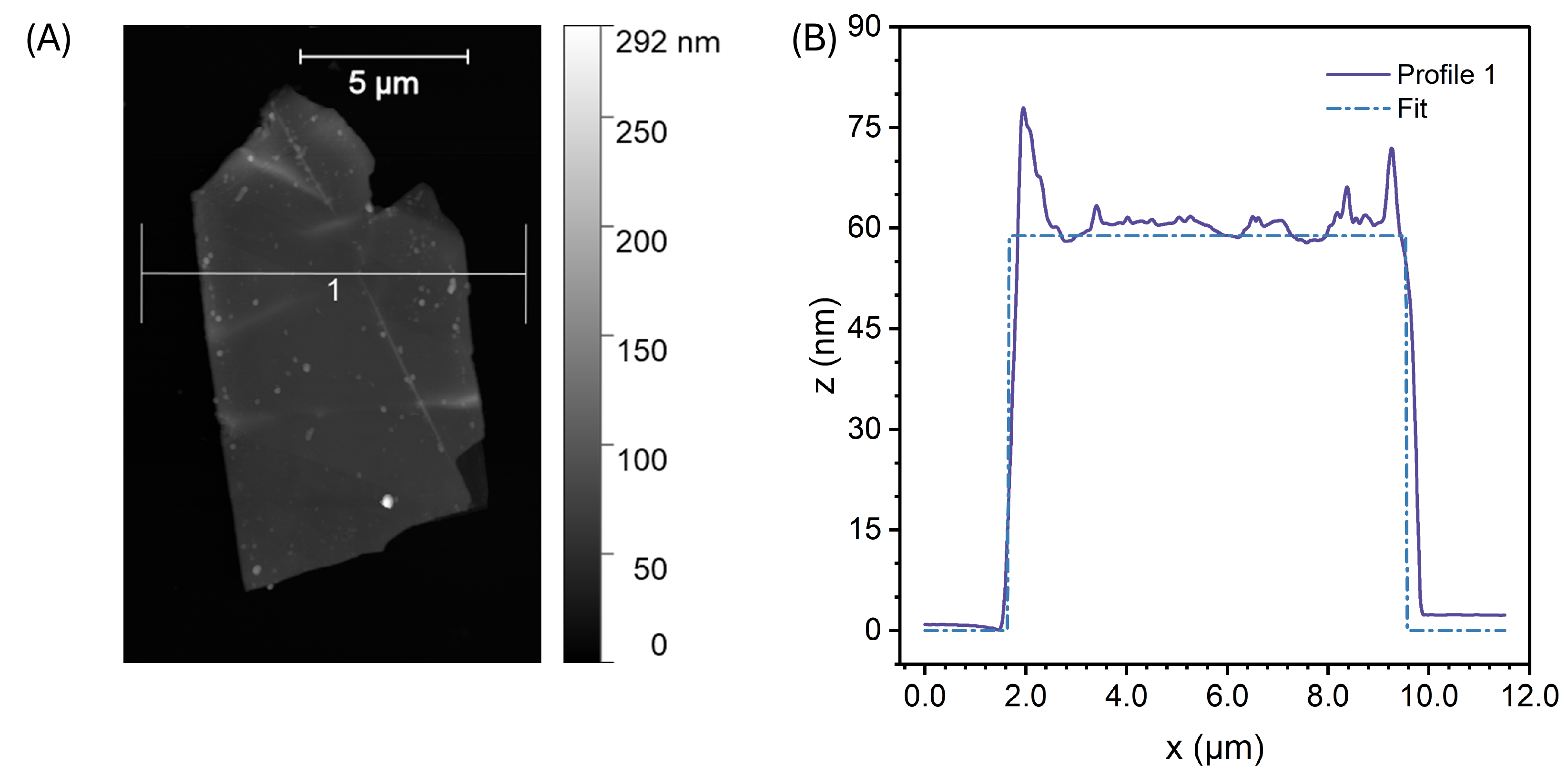}
    \caption{\textbf{Determination of flake thickness from AFM measurements.} (A) AFM topography of the exfoliated flake with the line profile indicated (profile width: 80 px). (B) Height profile averaged over the 80-px-wide line and fitted with a step-height model (dashed line). The fitted step height $h \approx 59~\mathrm{nm}$ is taken as the flake thickness.
    }
    \label{fig:AFM-PbTaSe2-thickness}
\end{figure}

\begin{figure}[H]
    \centering
    \includegraphics[width=\textwidth]{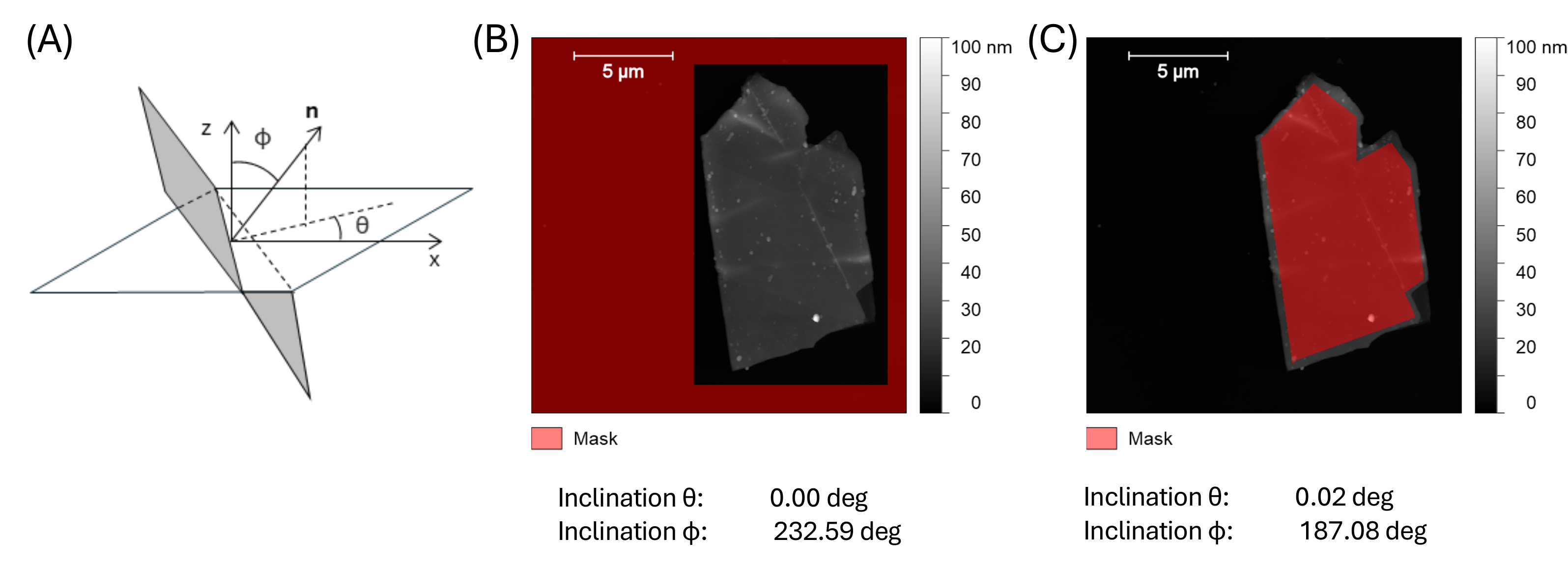}
    \caption{Assessment of flake tilt and geometrical definition of inclination angles. (A) Schematic illustrating the coordinate system and definition of the inclination angles $\theta$ and $\phi$ used in the plane-fitting analysis, where the $z$-axis corresponds to the surface normal. (B,C) Atomic force microscopy (AFM) topography of the SiO$_2$ substrate (B) and PbTaSe$_2$ flake (C). Planar fits were performed using masked regions (red) in Gwyddion to extract the local surface inclination. The resulting tilt angles are $\theta \approx 0.00^\circ$ for the substrate and $\theta \approx 0.02^\circ$ for the flake, indicating negligible out-of-plane tilt. This confirms that the surface normal is aligned with the crystallographic $c$-axis, justifying the assumption of $c \parallel z$ in the optical modeling.
    }
    \label{fig:AFM-plane inclination}
\end{figure}

\section{Isotropic-fit optical response of \texorpdfstring{PbTaSe$_2$}{PbTaSe2}} \label{sec:iso-fit}

\begin{figure}[H]
	\centering
	\includegraphics[width=\textwidth]{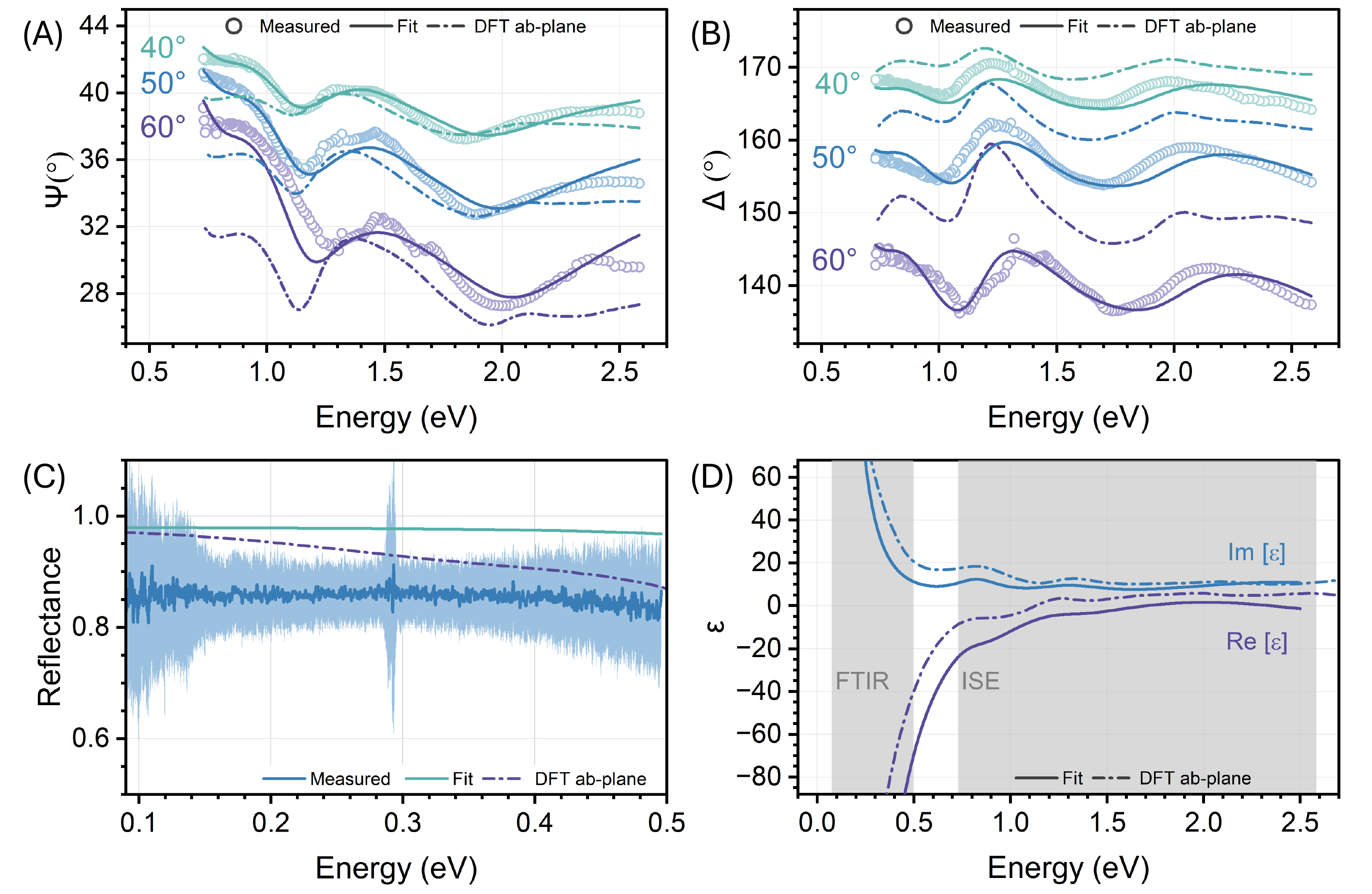}
    \caption{Isotropic-fit optical response of PbTaSe$_2$. (A,B) Ellipsometric $\Psi$ and $\Delta$ measured at angles of incidence of 40°, 50°, and 60° (symbols), together with the isotropic Drude–Lorentz (DL) fit (solid lines) consisting of three Lorentz oscillators and one Drude term (Table \ref{Isotropic_DL_Model}). The dashed lines show the response reconstructed from the DFT (ab-plane) dielectric function using the same multilayer optical model. (C) Normal-incidence FTIR reflectance at low photon energies. The blue solid line shows the measured FTIR data, with the shaded region indicating the experimental uncertainty ($\pm \sigma$). The green solid curve represents the isotropic Drude–Lorentz (DL) model obtained from fitting the imaging spectroscopic ellipsometry (ISE) data and subsequently extended into the IR region, capturing the expected Drude-like free-carrier response. The dashed curve shows the DFT (ab-plane) optical response over the same spectral range. (D) Real and imaginary parts of the effective isotropic dielectric function extracted from the DL model, compared with the DFT (ab-plane) response (dashed). The shaded region indicates the photon-energy range directly constrained by experimental data from FTIR and ellipsometry, while the unshaded region corresponds to model-based interpolation between the two measurement windows.}
	\label{fig:Isotropic-fit}
\end{figure}

Figures \ref{fig:Isotropic-fit}A and \ref{fig:Isotropic-fit}B present the ellipsometric angles $\Psi$ and $\Delta$ measured as a function of photon energy at angles of incidence of 40°, 50° and 60°, together with fits obtained using an effective isotropic Drude–Lorentz (DL) dielectric function which assumes $\overline{\overline{\varepsilon}}(E)\approx\varepsilon_{iso}(E)$:
\begin{equation}
\begin{aligned}
    \varepsilon_{iso}(E) &= \varepsilon_\infty + \varepsilon_{Drude} + \sum_{j=1}^{l}\varepsilon_{Lorentz,j} \\
    &= \varepsilon_\infty -\frac{A}{E^2+i{\Gamma}E} + \sum_{j=1}^{l} \frac{A_j}{E_{0,j}^2-E^2-i{\Gamma_j}E}. 
\end{aligned}
\end{equation}

The simultaneous and consistent agreement between the experiment and the model for both $\Psi$ and $\Delta$ at all angles demonstrates that an isotropic dielectric description is sufficient to capture the optical response of PbTaSe$_2$ over the measured spectral and angular ranges. The fit quality gets progressively worse at larger angles of incidence, as expected, since increasing angle enhances the sensitivity of the ellipsometric response to the out-of-plane dielectric component. Importantly, $\Psi$ and $\Delta$ encode complementary information: $\Psi$ is primarily sensitive to the amplitude of the reflection coefficients, while $\Delta$ probes their relative phase. The ability of a single isotropic dielectric function to reproduce both quantities simultaneously indicates that the dominant optical processes are well captured by a minimal set of physically meaningful contributions. Within this framework, the optical response is governed by a strong Drude term associated with free carriers and three Lorentz oscillators accounting for low energy and higher energy interband transitions (Table \ref{Isotropic_DL_Model}). 

As discussed in the main text, this isotropic description provides an effective baseline representation of the optical response, particularly useful for validating the multilayer optical model and capturing the dominant free-carrier and interband contributions.

\begin{table}
\centering
\caption{\textbf{Isotropic model: $\varepsilon_\infty=2.429$}}
\begin{tabular}{c|ccc}\hline
Oscillator &E$_0$ [eV] & A [eV$^2$] & $\Gamma$ [eV] \\
\hline
L1 & 2.550 & 45.191& 1.714\\
L2 & 1.319& 2.964& 0.480\\
L3 & 0.841& 2.117& 0.326\\
D1 & -& 21.444& 0.050\\\hline
\end{tabular}
\label{Isotropic_DL_Model}
\end{table}

This model $\varepsilon_{iso}(\omega)$ confirms that free carrier conductivity and low energy interband absorption dominate the optical response in this energy range and that these processes can be described without invoking directional anisotropy at the first level of analysis. Furthermore, agreement across multiple angles of incidence indicates that film thickness, interface quality, and substrate contributions are properly accounted for, establishing a reliable and internally consistent optical baseline.

To probe the optical response at low energy, the reflectance spectrum was measured using FTIR, Figure \ref{fig:Isotropic-fit}C. At normal incidence, the optical response probes an effective in-plane averaged dielectric function, making this measurement particularly well suited for independently validating the isotropic model extracted from ellipsometry. The reflectance is high (close to unity) and uniform over a wide spectral range, providing direct evidence of a strong free carrier response and confirming the metallic character of PbTaSe$_2$ at low energies. This behavior implies a large negative real part of the dielectric function at long wavelengths, which is a defining characteristic of metallic optical response. The FTIR reflectance therefore directly constrains the Drude contribution of the isotropic model and provides an independent experimental anchor for the low energy optical behavior inferred from ellipsometry.

Figure \ref{fig:Isotropic-fit}D presents $\mathrm{Re}[\varepsilon_{iso}]$ and $\mathrm{Im}[\varepsilon_{iso}]$ across the infrared to visible wavelengths. The shaded regions in Figure \ref{fig:Isotropic-fit}D indicate the photon energy ranges directly constrained by experimental FTIR and ellipsometry data, while the unshaded region corresponds to model based interpolation between the two measurement windows. This interpolation is physically constrained by the Drude–Lorentz formalism and ensures a continuous and Kramers–Kronig consistent dielectric response. The real part exhibits a large negative magnitude at low photon energies due to the dominant free carrier response and gradually increases toward zero as interband transitions become active. The imaginary part remains finite but moderate across the measured spectral range, reflecting non-negligible optical losses associated with both intraband scattering and interband absorption. This combination of strong metallic behavior and moderate dissipation places PbTaSe$_2$ in a regime that is favorable for supporting confined electromagnetic modes.

While this effective dielectric function accurately reproduces the measured optical response, it does not explicitly resolve the directional dependence of the dielectric tensor. As a result, phenomena that rely on the relative behavior of in-plane and out-of-plane components, such as dielectric sign inversions and hyperbolic dispersion, are addressed using the anisotropic model discussed in the main text.

\newpage

\section{Spin-orbit coupling effect on the dielectric function} \label{sec:SOC-dielectric-func}
The nodal-line electronic structure creates an extended set of degenerate valence and conduction states in momentum space. As a result, a larger number of low-energy interband transitions become optically allowed. This increases the low-energy contribution to $\varepsilon_2$ and produces pronounced spectral features near the onset of interband absorption. Through the Kramers-Kronig relations, these changes in $\mathrm{Im}[\varepsilon]$ also modify $\mathrm{Re}[\varepsilon]$. Because the nodal-line states are primarily composed of Ta $d$ and Pb/Se $p$ orbitals (Fig. \ref{fig:Crystal-Bands}) that couple strongly to in-plane electric fields, the effect is more pronounced for the in-plane component, $\varepsilon_{ab}$.

Including spin-orbit coupling (SOC) lifts band degeneracies and alters the low-energy band dispersion. This shifts the energies of interband transitions and redistributes optical oscillator strength. In the calculated dielectric functions, SOC therefore produces shifts and reshaping of spectral features in both $\mathrm{Re}[\varepsilon]$ and $\mathrm{Im}[\varepsilon]$, particularly at low photon energies where nodal-line-related transitions dominate. The impact of SOC is comparatively weaker for the out-of-plane component, $\varepsilon_c$, consistent with reduced interlayer transport and smaller optical matrix elements for electric fields polarized along the $c$ axis.

\begin{figure}[H]
	\centering
	\includegraphics[width=\textwidth]{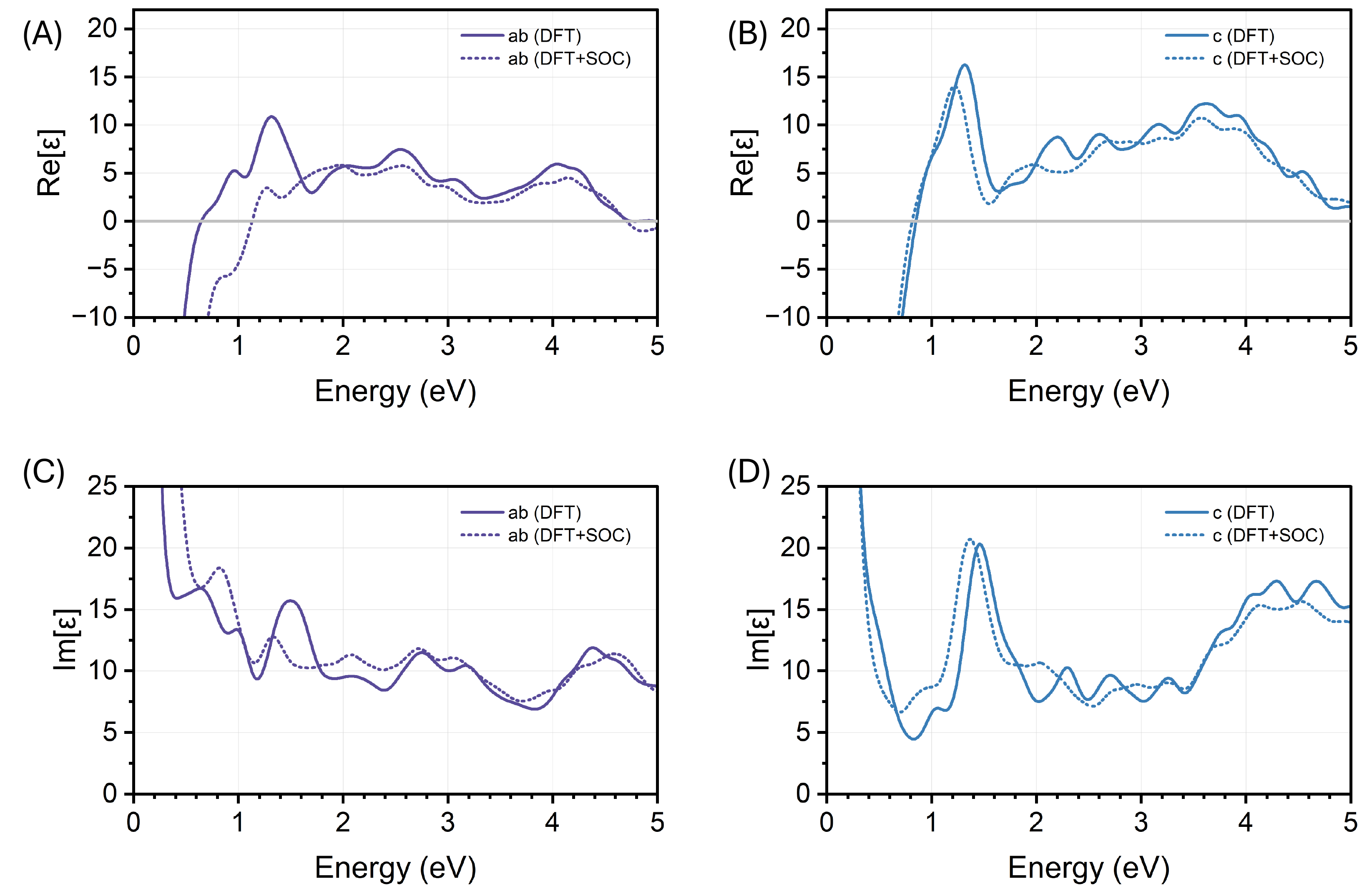}
    \caption{Effect of SOC on the PbTaSe$_2$ dielectric function. (A) In-plane $\mathrm{Re}[\varepsilon_{ab}]$.(B) In-plane $\mathrm{Im}[\varepsilon_{ab}]$. (\textbf{C}) Out-of-plane $\mathrm{Re}[\varepsilon_c]$. (D) Out-of-plane $\mathrm{Im}[\varepsilon_c]$.}
	\label{fig:SOC-effect}
\end{figure}

\newpage

\section{Calculation of the reflection coefficients} \label{sec:refl-coeff}

To model the ellipsometric response of the multilayer system, we calculate the complex reflection coefficients for \textit{p}- and \textit{s}-polarized light. The optical stack is modeled as a stratified multilayer structure with variation only along the $z$-direction. Each layer is assumed to be laterally homogeneous and infinite in the $xy$-plane, reducing the problem to one-dimensional propagation and enabling the use of recursive methods to compute the reflection coefficients. The multilayer structure is composed of air, the uniaxial PbTaSe$_2$ film, SiO$_2$, and the Si substrate. The PbTaSe$_2$ layer is described by a uniaxial dielectric tensor with in-plane and out-of-plane dielectric functions, $\varepsilon_{ab}$ and $\varepsilon_c$, respectively. The formalism employed here follows standard treatments of plane-wave propagation in stratified anisotropic media and multilayer ellipsometry~\cite{Fujiwara,Gold_2024,Gold_2025}.

For each wavelength and angle of incidence, the in-plane wavevector component is conserved across the multilayer and is given by
\begin{equation}
k_x = k_0 \sin\theta,
\end{equation}
where $k_0=2\pi/\lambda$ and $\theta$ is the angle of incidence in air.

For \textit{s}-polarized light, the out-of-plane wavevector component in layer $j$ is
\begin{equation}
k_{z,j}^{(s)} = \sqrt{\varepsilon_{ab,j}k_0^2-k_x^2},
\end{equation}

where $\varepsilon_{ab,j}$ is the in-plane component of the dielectric tensor of medium $j$, assuming a uniaxial medium.
For \textit{p}-polarized light, the out-of-plane-wave dispersion in the uniaxial medium gives
\begin{equation}
k_{z,j}^{(p)} =
\sqrt{
\varepsilon_{ab,j}k_0^2-
\frac{\varepsilon_{ab,j}}{\varepsilon_{c,j}}k_x^2
},
\end{equation}
where $\varepsilon_{c,j}$ is the out-of-plane component of the dielectric tensor of medium $j$. In isotropic layers, $\varepsilon_{ab,j}=\varepsilon_{c,j}$.

The single-interface Fresnel reflection coefficients between layers $j$ and $j+1$ are then calculated as
\begin{equation}
r_{j,j+1}^{(s)} =
\frac{k_{z,j}^{(s)}-k_{z,j+1}^{(s)}}
{k_{z,j}^{(s)}+k_{z,j+1}^{(s)}},
\end{equation}
and
\begin{equation}
r_{j,j+1}^{(p)} =
\frac{
\varepsilon_{ab,j}k_{z,j+1}^{(p)}
-
\varepsilon_{ab,j+1}k_{z,j}^{(p)}
}
{
\varepsilon_{ab,j}k_{z,j+1}^{(p)}
+
\varepsilon_{ab,j+1}k_{z,j}^{(p)}
}.
\end{equation}

Multiple reflections within the stack are included using a recursive multilayer expression. Starting from the bottom interface and moving upward, the effective reflection coefficient at interface $j-1$ is calculated as
\begin{equation}
\Gamma_{j-1}^{(\alpha)} =
\frac{
r_{j-1,j}^{(\alpha)}
+
\Gamma_{j}^{(\alpha)}
\exp\left(2ik_{z,j}^{(\alpha)}d_j\right)
}
{
1+
r_{j-1,j}^{(\alpha)}
\Gamma_{j}^{(\alpha)}
\exp\left(2ik_{z,j}^{(\alpha)}d_j\right)
},
\label{eq:reflection_recursion}
\end{equation}
where $\alpha=s,p$, $d_j$ is the thickness of layer $j$, and $\Gamma_j^{(\alpha)}$ is the effective reflection coefficient of the remaining layers below layer $j$.

The total reflection coefficients of the multilayer are therefore obtained as
\begin{equation}
r_s=\Gamma_0^{(s)}, \qquad r_p=\Gamma_0^{(p)}.
\end{equation}

The simulated ellipsometric ratio is then calculated from
\begin{equation}
\rho = \frac{r_p}{r_s}
      = \tan(\Psi)\exp\left[i(180^\circ-\Delta)\right].
\end{equation}

Thus, the modeled ellipsometric angles are obtained as
\begin{equation}
\Psi = \tan^{-1}|\rho|,
\end{equation}
and
\begin{equation}
\Delta = 180^\circ-\arg(\rho).
\end{equation}

\section{Confidence Interval Estimation}

Approximate 95\% confidence bands were estimated by Monte Carlo propagation of the parameter uncertainties derived from the covariance matrix of the nonlinear least-squares fit \cite{Morgan_Henrion_1990}. This approach accounts for parameter correlations and nonlinear propagation through the multilayer optical model. 

The fitting procedure minimizes the residual sum of squares between the experimental and modeled ellipsometric observables:
\begin{equation}
\chi^2 =
\sum_i
\left[
y_i^{\mathrm{exp}}
-
y_i^{\mathrm{model}}(\mathbf{p})
\right]^2,
\end{equation}
where $\mathbf{p}$ denotes the vector of fitted model parameters.

Near the optimum solution, the parameter covariance matrix was estimated from the Jacobian matrix of the residuals:
\begin{equation}
\mathbf{C}
=
\sigma^2
(\mathbf{J}^T\mathbf{J})^{-1},
\end{equation}
where $\mathbf{J}$ is the Jacobian matrix evaluated at the optimal solution and $\sigma^2$ is the variance of the residuals.

Parameter sets were then sampled from a multivariate normal distribution:
\begin{equation}
\mathbf{p}_{k}
\sim
\mathcal{N}
(\mathbf{p}_{\mathrm{opt}},\mathbf{C}),
\end{equation}
where $\mathbf{p}_{\mathrm{opt}}$ is the optimized parameter vector and $\mathbf{C}$ is the covariance matrix.

For each sampled parameter set, the dielectric functions and derived optical quantities were recalculated using the full multilayer optical model. The resulting distributions at each photon energy were used to estimate local confidence bands. The 95\% confidence intervals were defined from the 2.5 and 97.5 percentiles of the propagated distributions:
\begin{equation}
\mathrm{CI}_{95\%}
=
\left[
P_{2.5},
P_{97.5}
\right].
\end{equation}

\section{Reflection Measurement at 635 nm} 
\label{sec:refl-measurement}

\begin{figure}[H]
	\centering
	\includegraphics[width=\textwidth]{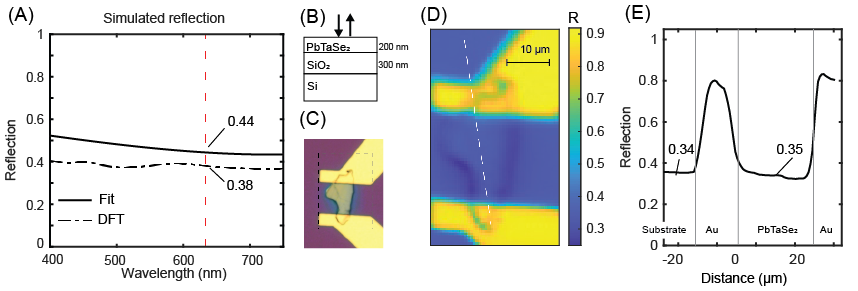}
    \caption{ Reflection at 635 nm. (A) Analytically modeled reflection spectrum of the PbTaSe$_2$ using the experimentally measured vs. the DFT calculated dielectric function, where (B) the flake is on an SiO$_2$/Si substrate. Reflection at 635 nm illumination was measured on (C) a PbTaSe$_2$ flake with gold electrodes. (D) A reflection map was collected with a scanning microscope. Reflection from the gold electrodes defines the absolute reflectance. The white dashed line corresponds to the reflection in (E). The reflectance of 0.35 on the flake has good agreement with the R = 0.38-0.44 range predicted by the DFT and ellipsometry-measured dielectric data.}
	\label{fig:635nm-reflection}
\end{figure}

To further probe the accuracy of the experimental and DFT-calculated permittivity values in the 1.7-2 eV spectral range, we mapped the reflection from a PbTaSe$_2$ flake at 635 nm = 1.95 eV. Reflection at normal incidence probes the in-plane behavior, and we analytically modeled the reflection for the two permittivity models in Figure \ref{fig:635nm-reflection}A-B using the recursion method in Section~\ref{sec:refl-coeff}. 

We then measured the reflection from the flake using a scanning reflection microscope (Fig. \ref{fig:635nm-reflection}C-E) which raster scanned a laser over the sample while reading out the reflected intensity with a photomultipler tube. 
At 635 nm illumination, the 50 nm Au electrodes on 10 nm Ti on the SiO$_2$/Si substrate have 91\% reflectivity, as calculated analytically. 
The absolute reflection of the sample was then calculated as $R = 0.91 {R_{\text{meas}}}/{R_{\text{Au-on-SiO}_2\text{/Si}}}$. The reflection from the bare SiO$_2$/Si substrate was measured as 0.34, which is in a good agreement with the simulated value of 0.35, verifying our measurement accuracy. The measured reflection from the flake is $\sim0.35$, in good agreement with the value predicted from the DFT and experimental dielectric functions of $~0.38-0.44$ (Fig.~\ref{fig:635nm-reflection}A).

\section{Previous reports of natural hyperbolic van der Waals materials} \label{sec:natural-hyperbolic-materials}

The data on natural hyperbolic van der Waals materials in Figure ~\ref{fig:Comparative} were extracted from the theoretical and experimental reports in Table~\ref{tab:natural-hyperbolic-materials}.

\begin{table}[h!]
\centering
\caption{Prior literature reports of natural hyperbolic van der Waals materials.}
\label{tab:natural-hyperbolic-materials}
\begin{tabular}{llll|ll}
\hline
\textbf{Material} & \textbf{Reference} & \textbf{Material} & \textbf{Reference} & \textbf{Material} & \textbf{Reference} \\
\hline
\multicolumn{4}{l}{\textit{DFT Data}} & \multicolumn{2}{l}{\textit{Experimental Data}} \\
\hline
1T-HfTe$_2$  & \cite{GNat_Comm_DFT} & 2H-MoS$_2$   & \cite{GNat_Comm_DFT} & Bi$_2$Te$_3$  & \cite{Bi2Se3_Bi2Te3_Esslinger} \\
1T-CoTe$_2$  & \cite{GNat_Comm_DFT} & PtS$_2$       & \cite{GNat_Comm_DFT} & Bi$_2$Se$_3$  & \cite{Bi2Se3_Bi2Te3_Esslinger} \\
1T-TiTe$_2$  & \cite{GNat_Comm_DFT} & 1T-ZrS$_2$   & \cite{GNat_Comm_DFT} & hBN           & \cite{hBN_Caldwell2014} \\
1T-RhTe$_2$  & \cite{GNat_Comm_DFT} & 2H-WS$_2$    & \cite{GNat_Comm_DFT} & Graphite      & \cite{graphite_exp} \\
1T-NiTe$_2$  & \cite{GNat_Comm_DFT} & 1T-SnS$_2$   & \cite{GNat_Comm_DFT} & 2H-NbSe$_2$  & \cite{NbSe2_TaSe2_exp} \\
1T-NbTe$_2$  & \cite{GNat_Comm_DFT} & PtSe$_2$      & \cite{GNat_Comm_DFT} & 2H-TaSe$_2$  & \cite{NbSe2_TaSe2_exp} \\
1T-IrTe$_2$  & \cite{GNat_Comm_DFT} & 2H-TaSe$_2$  & \cite{GNat_Comm_DFT} & 2H-WSe$_2$   & \cite{WSe2_exp} \\
1T-PdTe$_2$  & \cite{GNat_Comm_DFT} & 1T-TiSe$_2$  & \cite{GNat_Comm_DFT} & 1T-HfSe$_2$  & \cite{HfSe2_exp} \\
1T-PtTe$_2$  & \cite{GNat_Comm_DFT} & 2H-MoSe$_2$  & \cite{GNat_Comm_DFT} & 2H-MoS$_2$   & \cite{MoS2_exp} \\
1T-SiTe$_2$  & \cite{GNat_Comm_DFT} & 1T-ZrSe$_2$  & \cite{GNat_Comm_DFT} &  ZrSiSe$_2$   & \cite{Shao_2022} \\         
1T-ZrTe$_2$  & \cite{GNat_Comm_DFT} & 1T-HfSe$_2$  & \cite{GNat_Comm_DFT} &  CsV$_3$Sb$_5$   & \cite{Shiravi2024} \\  
1T-TaS$_2$   & \cite{GNat_Comm_DFT} & 2H-WSe$_2$   & \cite{GNat_Comm_DFT} &               & \\
2H-TaS$_2$   & \cite{GNat_Comm_DFT} & 1T-SnSe$_2$  & \cite{GNat_Comm_DFT} &               & \\
3R-NbS$_2$   & \cite{GNat_Comm_DFT} & 2H-NbSe$_2$  & \cite{GNat_Comm_DFT} &               & \\
2H-HfBrS     & \cite{GNat_Comm_DFT} & Graphite      & \cite{GNat_Comm_DFT} &               & \\
1T-TiS$_2$   & \cite{GNat_Comm_DFT} & 1T-PtSe$_2$  & \cite{GNat_Comm_DFT, PtSe2_Ghasemi2020} & & \\
1T-HfS$_2$   & \cite{GNat_Comm_DFT} &               &                       &               & \\
\hline
\end{tabular}
\end{table}

\clearpage
\printbibliography
\end{refsection}
\end{document}